\documentclass[twocolumn, pra, showpacs,superscriptaddress]{revtex4}
\usepackage{graphicx}
\usepackage{amsmath}
\usepackage{bm}
\usepackage{float}
\usepackage{color}

\begin{document}
\title{Quantum spin mixing in a binary mixture of spin-1 atomic condensates}
\author{Z. F. Xu}
\affiliation{State Key Laboratory of Low Dimensional Quantum
Physics, Department of Physics, Tsinghua University, Beijing 100084,
China}
\author{D. J. Wang}
\affiliation{Department of Physics,
The Chinese University of Hong Hong, Shatin, New Territories, Hong Kong, China}
\author{L. You}
\affiliation{State Key Laboratory of Low Dimensional Quantum Physics,
Department of Physics, Tsinghua University, Beijing 100084, China}

\date{\today}
\begin{abstract}
We study quantum spin mixing in a binary mixture of spin-1
condensates including coherent interspecies mixing process, using
the familiar spinor condensates of $^{87}$Rb and $^{23}$Na atoms in
the ground lower hyperfine $F=1$ manifolds as prototype examples.
Within the single spatial mode approximation for each of the two
spinor condensates, the mixing dynamics reduce to that
of three coupled nonlinear pendulums with clear physical
interpretations. Using suitably prepared initial states, it is
possible to determine the interspecies singlet-pairing as well as
spin-exchange interactions from the subsequent
mixing dynamics.
\end{abstract}


\maketitle

\section{Introduction}

A topical area in physics today concerns the control and manipulation
of the spinor degrees of freedom associated with electrons or atoms.
Two highly visible subfields attracting tremendous theoretical and experimental
interests are spintronics in condensed matter systems \cite{zutic2004} and
atomic spinor quantum gases \cite{ueda2010}.
The latter system become available due to the technical breakthrough of optical
trapping, which provides equal confinement for all atomic Zeeman components of a fixed $F$.
As a result, spin-related phenomena are exhibited and detected in cold atoms,
including various quantum phases \cite{ho1998,ohmi1998,law1998,ciobanu2000,koashi2000,ueda2002}
and quantum magnetism studies \cite{quantumM},
the observation of spin domain formation \cite{zhang2005l,sadler2006}, as well as the dynamics of
spin mixing \cite{law1998}, and spin squeezing \cite{nat1,nat2}, {\it etc}.

According to the formulation of atomic spinor condensates \cite{ho1998,ohmi1998,law1998,ciobanu2000,koashi2000,ueda2002},
the order parameter for a condensate in the hyperfine $F$ state is generally
described by a spinor of $2F+1$ components, strongly influenced by their
underline atom-atom interactions. Within the low energy limit of interests
to atomic quantum gases, when modeled by contact interactions,
atom-atom interactions are required to be invariant with respect
to both spatial and spin rotations, reflecting the nature of s-wave interactions.
Depending on the values of spin-dependent interaction parameters,
the ground state of a spinor condensate can be
ferromagnetic or anti-ferromagnetic (polar) for $F=1$ \cite{ho1998,ohmi1998,law1998},
while an additional cyclic phase appears when $F=2$ \cite{ciobanu2000,koashi2000,ueda2002}.
Higher spin cases are generally more complicated with limited experimental
access.

Law {\it et al.} pioneered the study of atomic spin mixing dynamics \cite{law1998}.
They first adopted numerical approach studying quantum spin mixing
in the absence of an external magnetic (B-) field \cite{law1998}.
Subsequent theoretical and experimental efforts have contributed to
the observation and control of the coherent quantum dynamics,
otherwise rarely visible in many body systems
\cite{barrett2001,schmaljohann2004,chang2004,kuwamoto2004,chang2005,kronjager2005,widera2006,kronjager2006,liu2009}.

In the semiclassical picture, using mean-field approximation
and adopting the single spatial mode approximation (SMA) \cite{law1998,yi2002},
coherent spin mixing dynamics in a spin-1 condensate is described by a nonrigid
pendulum, displaying periodic oscillations and resonance behavior
in an external B-field \cite{zhang2005a,romano2004}. This picture
proves to be widely popular with experimentalists
and led to many successes \cite{barrett2001,chang2004,chang2005,kronjager2005,widera2006,liu2009}.
Analogous efforts were applied to spin-2 condensates, for instance,
in the higher hyperfine manifold of the ground state $^{87}$Rb atoms \cite{schmaljohann2004,kuwamoto2004,widera2006,kronjager2006}.
An interesting application suggested by Saito {\it et al.} \cite{saito2005}
provides a practical method for determining the unknown spin coupling parameters
(polar or cyclic) relying on the mixing dynamics with suitably
prepared initial states.

Recently, several groups investigate intensively mixtures
of spinor condensates \cite{luo2007,xu2009,xu2010a,xu2010b,xu2010c,shi2010,xu2011},
whose properties are reasonably well understood,
both when an external B-field is absent or present.
As before for a single species spinor condensate,
semiclassical mean field approximations are adopted
and the full quantum approach is limited to atom number
dynamics in the restricted spatial modes of the condensates.
The ground state properties for the mixture, is to a large degree,
determined by the yet unknown interspecies spin exchange interaction.
If it is antiferromagnetic and is sufficiently strong, interesting
phases, such as highly fragmented ground states arise \cite{xu2010a,xu2010b}.
Additionally, there exists the so-called
broken-axisymmetry phase in the presence of an external B-field \cite{xu2010c}.
Within the degenerate internal state approximation \cite{stoof1988},
which considers atomic interaction potentials as
coming from contributions of potential curves
associated with the coupled electronic spins of the two valence electrons:
one for each type of atoms (taken as alkali atoms for simplicity)
 \cite{luo2007,weiss2003,pashov2005}.
The interspecies singlet-pairing interaction vanishes as
all interspecies interaction parameters
are determined by a total of only two scattering lengths for
the electronic singlet and triplet channels respectively.
This approximation provides a zeroth order estimates for
the $^{87}$Rb and $^{23}$Na atom mixture we study.
Experiences with spin exchange interactions within each species
show otherwise, i.e., the need for more atomic interaction parameters.

We therefore propose to develop analogous spin mixing dynamics
as in $F=2$ spinor condensates. We will calibrate the
interspecies singlet-pairing interactions with suitably
prepared initial states as in spin-2 condensates \cite{saito2005}.
Additionally, we find that inter-species
spin-exchange interaction can also be determined analogously.

\section{The model of a binary spin-1 condensate mixture}

The binary mixtures of spin-1 condensates have been discussed in
several earlier studies \cite{xu2009,xu2010a,xu2010b,xu2010c}.
In addition to the individual
Hamiltonian for each species of the two spinor condensates,
additional contact interactions exist between
the two species which can be decomposed into spin-independent
and spin-dependent terms as well, described by
$V_{\rm 12}(\vec{r}_1-\vec{r}_2)
=\frac{1}{2}(\alpha+\beta\mathbf{F}_1\cdot\mathbf{F}_2+\gamma\mathcal{P}_0)
\delta(\vec{r}_1-\vec{r}_2)$ \cite{xu2010a,xu2010b} with
appropriate interactions parameters
$\alpha$, $\beta$ and $\gamma$ \cite{xu2010a,xu2010b}.
Take spin-1 condensates of $^{87}$Rb and $^{23}$Na atoms as
examples, the total Hamiltonian is then given by
\begin{eqnarray}
  \hat{H}&=&\hat{H}_1+\hat{H}_2+\hat{H}_{12},\nonumber\\
  \hat{H}_1&=&\int d\mathbf{r}\,\left\{\hat{\Psi}^{\dag}_{m}
  \Big(-\frac{\hbar^2}{2M_1}\nabla^2+V_1^{o}-p_1 m+q_1 m^2 \Big)\hat{\Psi}_{m}\right. \nonumber\\
  &&\left.+\frac{\alpha_1}{2}
  \hat{\Psi}_i^{\dag}\hat{\Psi}_j^{\dag}\hat{\Psi}_j\hat{\Psi}_i
  +\frac{\beta_1}{2}\hat{\Psi}_i^{\dag}\hat{\Psi}_k^{\dag}
  \mathbf{F}_{1ij}\cdot\mathbf{F}_{1kl}\hat{\Psi}_l\hat{\Psi}_j\right\},\nonumber\\
  \hat{H}_{12}&=&\frac{1}{2}\int d\mathbf{r}\,\left\{\alpha
  \hat{\Psi}_i^{\dag}\hat{\Phi}_j^{\dag}\hat{\Phi}_j\hat{\Psi}_i
  +\beta\hat{\Psi}_i^{\dag}\hat{\Phi}_k^{\dag}\mathbf{F}_{1ij}\cdot
  \mathbf{F}_{2kl}\hat{\Phi}_l\hat{\Psi}_j
  \right.\nonumber\\
  &&\left.\qquad\qquad
  +\frac{1}{3}\gamma\,(-)^{i+j} \hat{\Psi}_i^{\dag}\hat{\Phi}_{-i}^{\dag}\hat{\Psi}_j\hat{\Phi}_{-j}\right\},
  \label{hamiltonian}
\end{eqnarray}
where $\hat{H}_1$ and $\hat{H}_2$ describe a single species system
of $^{87}$Rb and $^{23}$Na atoms respectively with the interspecies interaction
described by $H_{12}$.  $V_1^o$, $M_1$, $p_1$, and $q_1$ ($V_2^o$, $M_2$,
$p_2$, and $q_2$) respectively denote the optical trap,
atomic mass, linear, and quadratic Zeeman
shifts of a $^{87}$Rb ($^{23}$Na) atom. Both the nuclear spins and
the valence electron spins are the same for the two species.
In the subspace of hyperfine spin angular momentum $F=1$,
the linear Zeeman shifts for both $^{87}$Rb and $^{23}$Na atoms
are thus almost equal: $p_1\simeq p_2$ ($\equiv p$).
$\hat{\Psi}_i(\vec{r})$ ($\hat{\Phi}_i(\vec{r})$)
annihilate a $^{87}$Rb ($^{23}$Na) atom at the position $\vec{r}$.

The $F=1$ states for both $^{87}$Rb and $^{23}$Na atoms are well
studied, and their respective atomic collision parameters are known
precisely, quote their sources of respective $a_0$ and $a_2$, which
then gives $\alpha_{1/2}$ and $\beta_{1/2}$ \cite{Rb87,Na23}. While
a number of experimental and theoretical studies have previously
addressed collisions between $^{87}$Rb and $^{23}$Na atoms
\cite{weiss2003,pashov2005}, the most recent one by A. Pashov {\it
et al.} \cite{pashov2005} provides a well converged data set for
singlet and triplet scattering lengths of $a_s=70 (a_0)$ and
$a_t=109 (a_0)$. This can be used to predict the required set of
atomic intraspecies collision parameters $\alpha$, $\beta$, and
$\gamma$. What is certain concerns the value of spin exchange
interaction $\gamma$, it will be actually strong, instead of being
weak or vanishing. Perhaps we should consider using the real atomic
values for some of the calculations.

\begin{figure}[H]
\centering
\includegraphics[width=3.2in]{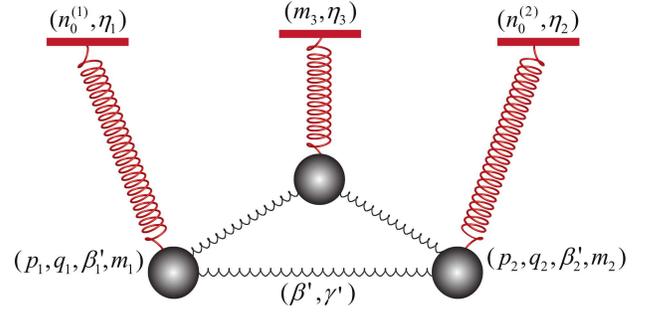}
\caption{(Color online). Schematic illustration of the three coupled nonlinear pendulums.}
\label{fig1}
\end{figure}

We now adopt the mean-field approximation and define for each condensate
species a mode function $\psi(\vec r)$/$\phi(\vec r)$,
justified by the fact the spin independent
density interaction terms are usually much stronger than
spin dependent interactions. We therefore take
$\Psi_i(\vec{r})\equiv\langle\hat{\Psi}_i(\vec{r})\rangle=\sqrt {n^{(1)}_j}\,e^{i\theta_j} \psi(\vec r)$ and
$\Phi_i(\vec{r})\equiv\langle\hat{\Phi}_i(\vec{r})\rangle=\sqrt {n^{(2)}_j}\,e^{i\varphi_j} \phi(\vec r)$.
The spin dynamics are then governed by
the spin-dependent energy functional
\begin{eqnarray}
  \mathcal{E}&=&\sum_{j=1,2}\mathcal{E}_j+\mathcal{E}_{12},\nonumber\\
  \mathcal{E}_{j}&=&-p_jm_j+q_j(n_j-n_0^{(j)})+\frac{1}{2}\beta'_jm_j^2
  \nonumber\\
  &&+\beta'_jn_0^{(j)}\Big[(n_j-n_0^{(j)})+\sqrt{(n_j-n_0^{(j)})^2-m_j^2}\cos\eta_j\Big],
  \nonumber\\
  \mathcal{E}_{12}&=&\frac{1}{2}\beta' m_1m_2
  +\frac{1}{6}\gamma'(n_1^{(1)}n_{-1}^{(2)}+n_0^{(1)}n_{0}^{(2)}+n_{-1}^{(1)}n_{1}^{(2)})
  \nonumber\\
  &&+\frac{1}{3}\gamma'\sqrt{n_1^{(1)}n_{-1}^{(1)}n_{1}^{(2)}n_{-1}^{(2)}}\cos\eta_3,
  \nonumber\\
  &&+(\beta'-\frac{1}{3}\gamma')\sqrt{n_0^{(1)}n_{-1}^{(1)}n_{1}^{(2)}n_0^{(2)}}\cos(\frac{\eta_1+\eta_2+\eta_3}{2})
  \nonumber\\
  &&+(\beta'-\frac{1}{3}\gamma')\sqrt{n_1^{(1)}n_0^{(1)}n_0^{(2)}n_{-1}^{(2)}}\cos(\frac{\eta_1+\eta_2-\eta_3}{2})
  \nonumber\\
  &&+\beta'\sqrt{n_0^{(1)}n_{-1}^{(1)}n_0^{(2)}n_{-1}^{(2)}}\cos(\frac{\eta_1-\eta_2+\eta_3}{2})
  \nonumber\\
  &&+\beta'\sqrt{n_1^{(1)}n_0^{(1)}n_{1}^{(2)}n_0^{(2)}}\cos(\frac{\eta_1-\eta_2-\eta_3}{2}),
  \label{energy}
\end{eqnarray}
where
$n_{1,2}=\sum_jn^{(1,2)}_j$, $m_{1,2}=n_1^{(1,2)}-n_{-1}^{(1,2)}$,
$\eta_1=\theta_1+\theta_{-1}-2\theta_0$,
$\eta_2=\varphi_1+\varphi_{-1}-2\varphi_0$,
and $\eta_3=\theta_{-1}-\theta_1+\varphi_1-\varphi_{-1}$.
The interaction parameters are now redefined to absorb the
relevant multipliers:
$\beta'_1=\beta_1\int|\psi(\vec r)|^4 d\vec r$,
$\beta'_2=\beta_2\int|\phi(\vec r)|^4 d\vec r$,
and $(\beta',\gamma')=(\beta,\gamma) \int|\psi(\vec r)|^2|\phi(\vec r)|^2 d\vec r$.
We note that $\int|\psi(\vec r)|^2 d\vec r=\int|\phi(\vec r)|^2 d\vec r=1$.
When the two species are immersible, the overlaps between
$\psi(\vec r)$ and $\phi(\vec r)$ is significantly reduced,
leading to diminished $\beta'$ and $\gamma'$, essentially reducing the
system to two stand-alone spin-1 condensates.

Although complicated in forms, the above Hamiltonian
gives rise to dynamics that can be interpreted simply
in terms of three coupled nonlinear pendulums,
with three pairs of canonical variables:
$(n_0^{(j)},\eta_j)$ and $(m_{3}=m_1-m_2,\eta_3)$. Their corresponding
equations of motion are given by
\begin{eqnarray}
  \dot{n}_0^{(j)}=-\frac{2}{\hbar}\frac{\partial \mathcal{E}}{\partial \eta_j},&\qquad&
  \dot{\eta}_j=\frac{2}{\hbar}\frac{\partial \mathcal{E}}{\partial n_0^{(j)}},
  \nonumber\\
  \dot{m}_{3}=-\frac{4}{\hbar}\frac{\partial \mathcal{E}}{\partial \eta_3},&\qquad&
  \dot{\eta}_3=\frac{4}{\hbar}\frac{\partial \mathcal{E}}{\partial m_{3}},
  \label{canonicaleq}
\end{eqnarray}
as illustrated schematically in Fig. \ref{fig1}.

\section{Determining the interspecies spin singlet-pairing interaction}

When discussing spin mixing in a spin-2 condensate,
Saito {\it et al.} \cite{saito2005} proposed to determine
the value of intra-species spin singlet-pairing interaction
by choosing an elementary process $0+0\leftrightarrow 2+(-2)$
which occurs only when the spin singlet-pairing interaction is non-vanishing.
With a suitable initial state of zero magnetization, the mixing
dynamics is governed by
coupled first-order ordinary differential equations,
which contain unknown parameters like singlet-pairing interactions
and quadratic Zeeman shifts. The analytic solutions can be compared with
the experimental measured dynamics to decide the unknowns.

Analogous approach can be taken to determine
the value of interspecies spin singlet-pairing interaction
for a binary mixture of spin-1 $^{87}$Rb and $^{23}$Na atom condensates,
making use of a different elementary collision process
$\Psi_1+\Phi_{-1}\leftrightarrow\Psi_{-1}+\Phi_1$
which appears only in the presence of the $\gamma'$ term.
With an initial state
\begin{eqnarray}
  \Psi/\psi=
  \left(
  \begin{array}{c}
    \sqrt{n_1^{(1)}}e^{i\theta_1} \\
    0 \\
    \sqrt{n_{-1}^{(1)}}e^{i\theta_{-1}}
  \end{array}
  \right),\quad
  \Phi_j/\phi=
  \left(
  \begin{array}{c}
    \sqrt{n_1^{(2)}}e^{i\varphi_1} \\
    0 \\
    \sqrt{n_{-1}^{(2)}}e^{i\varphi_{-1}}
  \end{array}
  \right),
  \label{initialstategamma'}
\end{eqnarray}
$\psi_0$ and $\phi_0$ re found to remain exactly zero
within the mean-field approximation,
unless dynamical instabilities exist.
If instabilities do occur, they can be suppressed by
tuning the quadratic Zeeman shifts $q_j$ to
a large negative value, for instance with off-resonant microwave field
\cite{gerbier2006,leslie2009}, or to a large positive value
with increased uniform B-field.
The processes $\Psi_1+\Psi_{-1}\leftrightarrow\Psi_0+\Psi_0$
and $\Phi_1+\Phi_{-1}\leftrightarrow\Phi_0+\Phi_0$
will then be suppressed, the populations
of the $M_F=0$ states remain at zero. In this case, the spin mixing dynamics
of Eq. (\ref{canonicaleq}) reduce to that of a single pair,
which takes the form,
\begin{eqnarray}
  \dot{m}_{3}&=&\frac{\gamma'}{12\hbar}\sqrt{[4n_1^2-(m+m_{3})^2][4n_2^2-(m-m_{3})^2]}\nonumber\\
  &&\times\sin\eta_3,
  \nonumber\\
  \dot{\eta}_3&=&\frac{\beta'_1-\beta'_2}{\hbar}m+\frac{\beta'_1+\beta'_2-\beta'+\gamma'/6}{\hbar}m_{3}
  \nonumber\\
  &-&\frac{\gamma'}{6\hbar}
  \frac{2(n_1^2+n_2^2)m_{3}-2(n_1^2-n_2^2)m+m^2m_{3}-m_{3}^3}
  {\sqrt{[4n_1^2-(m+m_{3})^2][4n_2^2-(m-m_{3})^2]}}\nonumber\\
  &&\times\cos\eta_3,
  \label{canonicaleqgamma'}
\end{eqnarray}
and is described by a simpler energy functional
\begin{eqnarray}
  \mathcal{E}&=&\frac{\beta'_1+\beta'_2-\beta'+\gamma'/6}{8}m_{3}^2
  +\frac{\beta'_1-\beta'_2}{4}mm_{3}
  \nonumber\\
  &+&\frac{\gamma'}{48}\sqrt{[4n_1^2-(m+m_{3})^2][4n_2^2-(m-m_{3})^2]}\cos\eta_3, \hskip 12pt
  \quad
  \label{energygamma'}
\end{eqnarray}
after neglecting a constant term $-pm+(\beta'_1+\beta'_2+\beta'-\gamma'/6)m^2/8+\gamma' n_1n_2/12+q_1n_1+q_2n_2$.
Substituting Eq. (\ref{energygamma'}) into Eq. (\ref{canonicaleqgamma'}), we find
\begin{eqnarray}
  &&(\dot{m}_{3})^2=\left(\frac{\gamma'}{12\hbar}\right)^2
  [4n_1^2-(m+m_{3})^2][4n_2^2-(m-m_{3})^2]
  \nonumber\\
  &&-\frac{16}{\hbar^2}\Big[\mathcal{E}-\frac{\beta'_1+\beta'_2-\beta'+\gamma'/6}{8}m_{3}^2
  -\frac{\beta'_1-\beta'_2}{4}mm_{3}\Big]^2,
  \label{canonicaleqgamma'2}
\end{eqnarray}
which can be integrated following the procedure of
Ref. \cite{zhang2005a} by solving for the equation
$\dot{m}_{3}=0$, keeping the
interspecies spin-dependent interaction parameters $\beta'$ and $\gamma'$
as unknown.

From the Eq. (\ref{canonicaleqgamma'}) and assuming an initial state with $\sin\eta_3>0$,
we infer $\gamma'>0$ if $m_{3}$ increases during the initial short time period of the spin
mixing dynamics, and $\gamma'<0$ if it decreases.
To determine $\gamma'$, we prepare an initial state
with $\eta_3=\pi/2$ and $m_1=m_2=0$, which leads to
$\mathcal{E}=0$ in the Eq. (\ref{energygamma'}), and
\begin{eqnarray}
  (\dot{m}_{3})^2=\frac{\gamma'^2}{144\hbar^2}
  \Big[(m_{3}^2-4n_1^2)(m_{3}^2-4n_2^2)
  -\mathcal{C}^2m_3^4\Big],
  \label{dottm}
\end{eqnarray}
with $\mathcal{C}=|6(\beta'_1+\beta'_2-\beta')/\gamma'+1|$.
If $\mathcal{C}<1$, $\dot{m}_3=0$ gives
four roots $-x_2$, $-x_1$, $x_1$, and $x_2$, where
$x_{1/2}=\sqrt{2\big(n_1^2+n_2^2\mp\sqrt{(n_1^2-n_2^2)^2+4\mathcal{C}^2n_1^2n_2^2}\big)/(1-\mathcal{C}^2)}$.
For $\mathcal{C}\ge1$, however, only two solutions $-x_1$ and $x_1$ exist.
The solution for the mixing dynamics is expressed
in terms of the Jacobian elliptic functions sn(.) and cn(.) as
\begin{eqnarray}
  m_{3}(t)&=&x_1\ {\rm sn}
  \left(\frac{x_2\gamma' t\sqrt{1-\mathcal{C}^2}}{12\hbar},
  \frac{x_1}{x_2}
  \right), \text{for $\mathcal{C}\le1$},\nonumber\\
  m_{3}(t)&=&x_1\ {\rm cn}
  \left(K\Big(\frac{x_1}{\sqrt{x_1^2+x_3^2}}\Big)
  -\frac{\gamma' t\sqrt{(x_1^2+x_3^2)(\mathcal{C}^2-1})}{12\hbar},\right.
  \nonumber\\
  &&\qquad\left.\frac{x_1}{\sqrt{x_1^2+x_3^2}}
  \right),
  \text{for $\mathcal{C}\ge1$},
  \label{tmgamma'}
\end{eqnarray}
where $K(.)$ is the complete elliptic integral of the first kind,
and $x_3=\sqrt{2\big(n_1^2+n_2^2+\sqrt{(n_1^2-n_2^2)^2+4\mathcal{C}^2n_1^2n_2^2}\big)/(\mathcal{C}^2-1)}$.

\begin{figure}[tbp]
\centering
\includegraphics[width=3.0in]{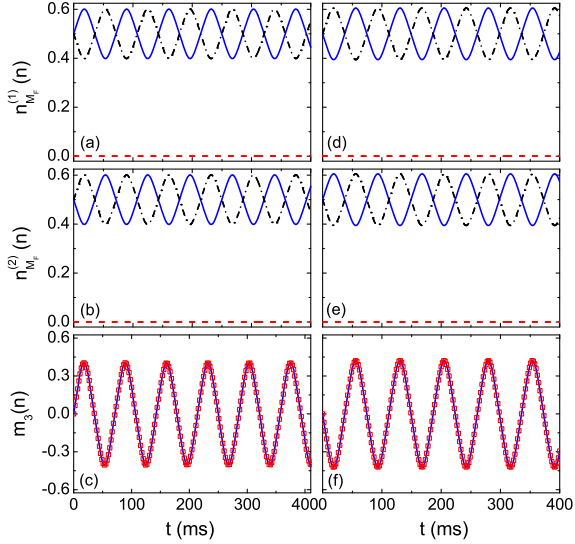}
\caption{(Color online). Time dependent populations of
each spin components. In the left panels of (a)-(c),
the interspecies interaction parameters used are
$\beta'=5|\beta'_1|$ and $\gamma'=2|\beta'_1|$.
For the panels of (d)-(f), $\beta'=5|\beta'_1|$ and
$\gamma'=-2|\beta'_1|$ are used.
(a) For the $^{87}$Rb condensate, where the solid blue line,
dashed red line, and dotted-dash black line represent
the $M_F=1,0,-1$ components, respectively.
(b) As in (a), but for the $^{23}$Na condensate.
(c) Time dependent $m_3$ with solid blue line and red square symbols
denote numerical and analytical solutions respectively.
(d) As (a), but with $\gamma'=-2|\beta'_1|$. (e) As in (b), but
with $\gamma'=-2|\beta'_1|$. (f) As in (c), but with $\gamma'=-2|\beta'_1|$. }
\label{fig2}
\end{figure}

The stability of the above dynamics are confirmed with numerical solutions,
taking the initial state as
$\Psi_j/\psi=\sqrt{n_1}(1,0,1)^T/\sqrt{2}$, $\Phi_j/\phi=\sqrt{n_2}(1,0,-i)^T/\sqrt{2}$,
assuming $n_1=n_2=n$ and $n=2\times10^4$.
We further choose $\beta'_1/\hbar=-22.4893\times10^{-4}\,\rm Hz $ and
$\beta'_2/\hbar=303.816\times10^{-4}\,\rm Hz$.
A noise level at $10^{-5}$ in the population of $M_F=0$
spin states of both species is also included.
The B-field is set as large enough to suppress the intraspecies
spin-exchange process with the quadratic Zeeman shifts
satisfying $q_1=40 |\beta'_1|n$ and $q_2=q_1\Delta E_1/\Delta E_2$,
where $\Delta E_1$ and $\Delta E_2$ are the hyperfine splittings
of $^{87}$Rb and $^{23}$Na atoms respectively.
Figure \ref{fig2} illustrates our numerical results.
In Fig. \ref{fig2}(a-c), $\beta'=5|\beta'_1|$ and  $\gamma'=2|\beta'_1|$ are used,
while $\beta'=5|\beta'_1|$ and $\gamma'=-2|\beta'_1|$ are used instead for Fig. \ref{fig2} (d-f).
The time evolution for each condensate species are shown
in Fig. \ref{fig2}(a,d) and (b,e) respectively for $^{87}$Rb and $^{23}$Na atoms.
The evolutions for $m_3$ are shown in Fig. \ref{fig2}(c) and (f),
indeed they confirm our predictions based on the insights gained
from analytical solutions that $m_3$ increases/decreases
at the beginning when $\gamma'>0$/$\gamma'<0$.
The numerical simulations denoted by solid blue lines agree well with
analytical solutions of Eq. (\ref{tmgamma'}) denoted by red square symbols.
We further note that
$\dot{n}^{(1)}_1=-\dot{n}^{(1)}_{-1}=-\dot{n}^{(2)}_1=\dot{n}^{(1)}_{2}=
\dot{m}_3/4$ with the initial state used in this case.
As a result, we can determine the sign of $\gamma'$
from the populations of arbitrary spin components and species.

Using Eq. (\ref{tmgamma'}),  we can then proceed to
determine the value of $\gamma'$ if
$\mathcal{C}^2=x_1^2-4n_1^2-4n_2^2+16n_1^2n_2^2/x_1^2$ is first determined
from the oscillation amplitude of $m_3$.
Afterwards, $\beta'$ becomes partially determine to within the
following two choices
\begin{eqnarray}
  \beta'_\mp&=&\beta'_1+\beta'_2\mp(\mathcal{C}\mp1)\gamma'/6.
  \label{beta'}
\end{eqnarray}

\section{Determining the interspecies spin-exchange interaction}

In the previous section, a scheme is proposed capable of
determining the interspecies singlet-pairing interaction parameter $\gamma'$
following spin mixing dynamics from a suitably chosen initial state.
The interspecies spin-exchange interaction parameter $\beta'$,
which is partially determined at the same time,
will become fully determined with the dynamics discussed in this section.

Equation (\ref{energy}) gives four relevant elementary
spin-exchange processes:
$\Psi_0+\Phi_0\leftrightarrow\Psi_{-1}+\Phi_{1}$,
$\Psi_0+\Phi_0\leftrightarrow\Psi_{1}+\Phi_{-1}$,
$\Psi_{-1}+\Phi_0\leftrightarrow\Psi_{0}+\Phi_{-1}$,
and
$\Psi_1+\Phi_0\leftrightarrow\Psi_{0}+\Phi_{1}$,
which can be used to determine the interspecies spin-exchange interaction.
The first two processes involve both $\beta'$ and $\gamma'$ terms
of the Hamiltonian in Eq. (\ref{hamiltonian}),
while the last two processes are solely induced by spin-exchange
interactions.

The above four individual processes become independent
if all other possible collision channels are suppressed.
For example, to observe the mixing due to
$\Psi_0+\Phi_0\leftrightarrow\Psi_{-1}+\Phi_{1}$, $\Psi_1=\Phi_{-1}=0$
needs to be ensured at all times. A plausible scenario can again
employ increased quadratic Zeeman shifts
$q_1$ and $q_2$. As long as the energy difference between the final
state and the initial state increases, the intraspecies spin-exchange
process $\Psi_1+\Psi_{-1}\leftrightarrow\Psi_0+\Psi_0$ and
$\Phi_1+\Phi_{-1}\leftrightarrow\Phi_0+\Phi_0$ are suppressed.
They help to maintain close to zero populations during time evolution
in the corresponding spin state, if an initial state with $\Psi_1=\Phi_{-1}=0$ is used.

In the following we will describe the isolation of the process
$\Psi_1+\Phi_0\leftrightarrow\Psi_{0}+\Phi_{1}$ as an example
to determine the interspecies spin-exchange interaction.
An initial state
\begin{eqnarray}
  \Psi/\psi=
  \left(
  \begin{array}{c}
    \sqrt{n_1^{(1)}}e^{i\theta_1} \\
    \sqrt{n_{0}^{(1)}}e^{i\theta_{0}} \\
    0
  \end{array}
  \right),\quad
  \Phi/\phi=
  \left(
  \begin{array}{c}
    \sqrt{n_1^{(2)}}e^{i\varphi_1} \\
    \sqrt{n_{0}^{(2)}}e^{i\varphi_{0}} \\
    0
  \end{array}
  \right),
  \label{initialstatebeta}
\end{eqnarray}
is assumed, together with a sufficiently strong uniform external magnetic field
to guarantee $\Psi_{-1}=\Phi_{-1}=0$.
Since interspecies spin mixing is only induced by the same $\beta'$ term,
and energy conservation, the spin mixing dynamics is then governed by
the evolution of $m_3$
\begin{eqnarray}
  \dot{m}_{3}&=&-\frac{\beta'}{2\hbar}\sqrt{(m^2-m_{3}^2)(2n_1-m-m_{3})(2n_2-m+m_{3})}
  \nonumber\\
  &&\times \sin\frac{\eta_1-\eta_2-\eta_3}{2},
  \label{cannonicaleqbeta}
\end{eqnarray}
which can be derived from the Eq. (\ref{canonicaleq}),
and the associated energy functional
\begin{eqnarray}
  \mathcal{E}&=&-\frac{\beta'_1+\beta'_2+\beta'}{8}m_{3}^2
  -\frac{\beta'_1-\beta'_2}{4}mm_{3}
  +\frac{\beta'_1n_1+q_1}{2}m_{3}
  \nonumber\\
  &-&\frac{\beta'_2n_2+q_2}{2}m_{3}
  +\frac{\gamma'}{24}(2n_1-m-m_{3})(2n_2-m+m_{3})
  \nonumber\\
  &+&\frac{\beta'}{4}\sqrt{(m^2-m_{3}^2)(2n_1-m-m_{3})(2n_2-m+m_{3})}
  \nonumber\\
  &&\times\cos\frac{\eta_1-\eta_2-\eta_3}{2},
  \label{energybeta}
\end{eqnarray}
after neglecting a constant term
$-pm-(\beta'_1+\beta'_2-\beta')m^2/8+(\beta'_1n_1+\beta'_2n_2+q_1+q_2)m/2$.
Furthermore we can rewrite Eq. (\ref{cannonicaleqbeta}) as
\begin{eqnarray}
  (\dot{m}_{3})^2&=&\frac{\beta'^2}{4\hbar^2}(m^2-m_{3}^2)(2n_1-m-m_{3})(2n_2-m+m_{3})
  \nonumber\\
  &-&\frac{4}{\hbar^2}\Big[\mathcal{E}
  +\frac{\beta'_1+\beta'_2+\beta'}{8}m_{3}^2
  +\frac{\beta'_1-\beta'_2}{4}mm_{3}
  \nonumber\\
  &&\quad\quad -\frac{\gamma'}{24}(2n_1-m-m_{3})(2n_2-m+m_{3})
  \nonumber\\
  &&\quad\quad-\frac{\beta'_1n_1+q_1}{2}m_{3}
  +\frac{\beta'_2n_2+q_2}{2}m_{3}\Big]^2.
  \label{canonicaleqbeta2}
\end{eqnarray}

The procedure to fully determine $\beta'$ goes as follows.
First we infer the sign of $\beta'$ from the initial stage of the
time evolution for $m_3$, as in the earlier section on
determining the sign of $\gamma'$.
For an initial state with $\Psi/\psi=\sqrt{n_1}(1,1,0)^T/\sqrt{2}$
and $\Phi/\phi=\sqrt{n_2}(1,-i,0)^T/\sqrt{2}$, where
$(\eta_1-\eta_2-\eta_3)/2=\theta_1-\theta_0-\varphi_1+\varphi_0=-\pi/2$,
we confirm $\beta'>0$ ($\beta'<0$) if $m_3$ initially increases (decreases).
The actual value of $\beta'$ is determined by comparing the
analytic or numerical solutions using the two choices
of $\beta'$ from the Eq. (\ref{beta'}) to experimental measurements.
Again we assume $n_1=n_2=n=2\times10^4$, $\beta'_1/\hbar=-22.4893\times10^{-4}\,\rm Hz$,
$\beta'_2/\hbar=303.816\times10^{-4}\,\rm Hz$,
$\beta'=5|\beta'_1|$, $\gamma'=2|\beta'_1|$,
$q_1=30 |\beta'_1|n$, and $q_2=q_1\Delta E_2/\Delta E_1$, with the
analytic solution for $m_3$
\begin{eqnarray}
  m_3(t)&=&\frac{x_1(x_2-x_4)+(x_1-x_2)x_4x^2}{(x_2-x_4)+(x_1-x_2)x^2},\nonumber\\
  x&=&{\rm sn}\left(d_4-t\sqrt{d_3}/d_2, d_1)\right),
  \label{tmbeta}
\end{eqnarray}
where $x_{j=1,2,3,4}$ are the four roots of $\dot{m}_3=0$,
arranged in descending order $x_1>0>x_2>x_3>x_4$, and
$d_1=\sqrt{(x_1-x_2)(x_3-x_4)/(x_1-x_3)/(x_2-x_4)}$,
$d_2=2/\sqrt{(x_1-x_3)(x_2-x_4)}$, $d_3=[4\beta'^2-(\beta'_1+\beta'_2+\beta'+\gamma'/3)^2]/16\hbar^2$,
and
$d_4={\rm F}(\arcsin\sqrt{-(x_2-x_4)x_1/(x_1-x_2)/x_4},d_1)$,
with F(.) the elliptic integral of the first kind.

\begin{figure}[tbp]
\centering
\includegraphics[width=3.2in]{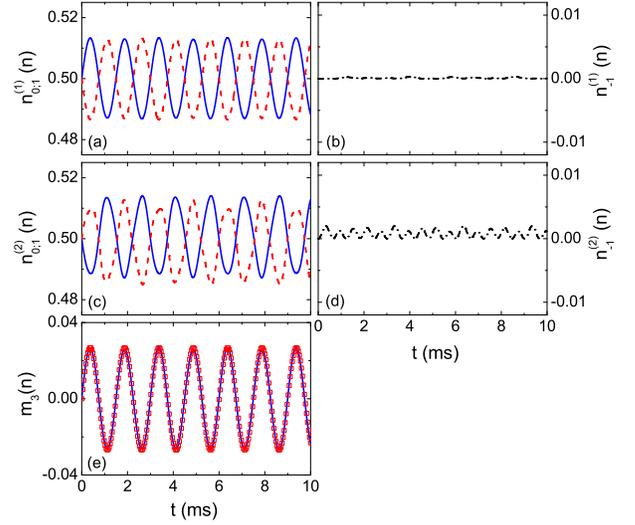}
\caption{(Color online). Population dynamics for all spin components,
with the interspecies interaction parameters
$\beta'=5|\beta'_1|$ and $\gamma'=2|\beta'_1|$,
and the quadratic Zeeman shifts  $q_1=30 |\beta'_1|n$
and $q_2=q_1\Delta E_2/\Delta E_1$.
(a) For the $^{87}$Rb condensate, where the solid blue line and
dashed red line represent the $M_F=1,0$ components, respectively.
(b) As in (a), but for the $M_F=-1$ component in dotted-dash black line.
(c)/(d) corresponds to that in (a)/(b) respectively,
but for the $^{23}$Na condensate.
(e) Time evolution of $m_3$, where solid blue line and red square symbols
denote numerical and analytical solutions respectively.}
\label{fig3}
\end{figure}

Figure \ref{fig3} show population evolutions for
all spin components. Due to the large but unequal
quadratic Zeeman shifts $q_1$ and $q_2$,
the suppression of intraspecies spin mixing dynamics
leads to a suppressed amplitude for interspecies spin-exchange
dynamics. As a result, the quadratic Zeeman shifts cannot
be tuned too large, otherwise they cause nonzero populations in the $M_F=-1$ spin component
especially for the $^{23}$Na atoms as illustrated in Fig. \ref{fig3}(d).

The other three elementary channels can also be employed
to determine the interspecies spin-exchange interaction. Among them,
two are capable of determining the combined parameter
$\beta'-\gamma'/3$, which can be further aided by a determination
of the sign of $\beta'-\gamma'/3$.

Before conclusion, we hope to stress that the special mixture illustrated
in this study involves a spin-1 condensate with ferromagnetic interaction ($^{87}$Rb)
and a polar spin-1 condensate ($^{23}$Na) with antiferromagnetic interaction.
More generally the procedures we suggest for determine the interspecies
interaction parameters remain applicable for mixtures with two spin-1
ferromagnetic condensates or two antiferromagnetic condensates.

\section{Conclusion}

We discuss coherent spin mixing dynamics for a binary mixture of
spin-1 condensates. Under the mean field approximation,
the dynamics reduce to three coupled nonlinear pendulums,
one for each spin-1 condensate as understood previously for stand-alone
spin-1 condensate \cite{zhang2005a}, and a third one for the
difference in magnetization between the two species.
By tuning quadratic Zeeman shifts to large enough values,
they can suppress intraspecies spin mixing dynamics,
which results in a pure interspecies spin mixing dynamics.
Using suitably prepared initial states with zero populations
in the $M_F=0$ states for both species, we can determine
the value of the interspecies singlet-pairing interaction
by comparing the analytic formula to experimental measurements,
and at the same time we can partially determine the value of the
interspecies spin-exchange interaction parameter $\beta'$.
Next, using an alternative initial state with zero populations in the
$M_F=-1$ states of both species, and using the two possible values
for $\beta'$ partially determined above, we can numerically
or analytically solve the dynamics and compare them with
experimental results to determine the
correct value of $\beta'$.

\section{Acknowledgements}

This work is supported by NSF of China under Grant No. 11004116,
No. 91121005, NKBRSF of China, and the research program 2010THZO of
Tsinghua University. D. W. is supported by Hong Kong RGC CUHK
403111.


\begin{thebibliography}{10}

\bibitem{zutic2004}
  I. \v{Z}uti\'c, J. Fabian, S. Das Sarma, Rev. Mod. Phys. \textbf{76}, 323 (2004).

\bibitem{ueda2010}
  M. Ueda and Y. Kawaguchi, e-print arXiv: 1001.2072.

\bibitem{ho1998}
  Tin-Lun Ho, Phys. Rev. Lett. \textbf{81}, 742 (1998).

\bibitem{ohmi1998}
  T. Ohmi and K. Machida, J. Phys. Soc. Jpn. \textbf{67}, 1822
  (1998).

\bibitem{law1998}
  C. K. Law, H. Pu, and N. P. Bigelow, Phys. Rev. Lett.
  \textbf{81}, 5257 (1998).

\bibitem{koashi2000}
  Masato Koashi and Masahito Ueda, Phys. Rev. Lett. \textbf{84}, 1066 (2000).

\bibitem{ciobanu2000}
  C. V. Ciobanu, S.-K.Yip, and Tin-Lun Ho, Phys. Rev. A \textbf{61}, 033607 (2000).

\bibitem{ueda2002}
  M. Ueda and M. Koashi, Phys. Rev. A \textbf{65}, 063602 (2002).

\bibitem{quantumM}
A. M. Rey, V. Gritsev, I. Bloch, E. Demler, and M. D. Lukin,
Phys. Rev. Lett. {\bf 99}, 140601 (2007).

\bibitem{zhang2005l}
  Wenxian Zhang, D. L. Zhou, M.-S. Chang, M. S. Chapman, and L. You,
  Phys. Rev. Lett. \textbf{95}, 180403 (2005).

\bibitem{sadler2006}
  L. E. Sadler, J. M. Higbie, S. R. Leslie, M. Vengalattore, and D. M. Stamper-Kurn,
  Nature {\bf 443}, 312 (2006).

\bibitem{nat1}C. Gross, T. Zibold, E. Nicklas, J. Est\`{e}ve, and M. K. Oberthaler,
Nature \textbf{464}, 1165 (2010).

\bibitem{nat2} M. F. Riedel, P. B\"{o}hi, Y. Li, T. W. H\"{a}nsch, A. Sinatra, and P. Treutlein,
Nature \textbf{464}, 1170 (2010).

\bibitem{barrett2001}
  M. D. Barrett, J. A. Sauer, and M. S. Chapman, Phys. Rev. Lett.
  \textbf{87}, 010404 (2001).

\bibitem{chang2004}
  M.-S. Chang, C. D. Hamley, M. D. Barrett, J. A. Sauer, K. M. Fortier, W. Zhang, L. You,
  and M. S. Chapman, Phys. Rev. Lett. \textbf{92}, 140403 (2004).

\bibitem{chang2005}
  M.-S. Chang, Q. Qin, W. Zhang, L. You, M. S. Chapman,
  Nature Physics \textbf{1}, 111 (2005).

\bibitem{kronjager2005}
  J. Kronj\"ager, C. Becker, M. Brinkmann, R. Walser, P. Navez, K. Bongs, and K. Sengstock,
  Phys. Rev. A \textbf{72}, 063619 (2005).

\bibitem{liu2009}
  Y. Liu, S. Jung, S. E. Maxwell, L. D. Turner, E. Tiesinga, and P. D. Lett,
  Phys. Rev. Lett. \textbf{102}, 125301 (2009).

\bibitem{widera2006}
  A. Widera, F. Gerbier, S. F\"olling, T. Gericke, O. Mandel and I. Bloch,
  New J. Phys. \textbf{8}, 152 (2006).

\bibitem{schmaljohann2004}
  H. Schmaljohann, M. Erhard, J. Kronj\"ager, M. Kottke,
  S. van Staa, L. Cacciapuoti, J. J. Arlt, K. Bongs, and K. Sengstock,
  Phys. Rev. Let. \textbf{92}, 040402 (2004).

\bibitem{kuwamoto2004}
  T. Kuwamoto, K. Araki, T. Eno, and T. Hirano, Phys. Rev. A \textbf{69},
  063604 (2004).

\bibitem{kronjager2006}
  J. Kronj\"ager, C. Becker, P. Navez, K. Bongs, and K. Sengstock,
  Phys. Rev. Lett. \textbf{97}, 110404 (2006).

\bibitem{yi2002}
  S. Yi, \"O. E. M\"ustecapl{\i}o\u{g}lu, C. P. Sun, and L. You, Phys. Rev. A
  \textbf{66}, 011601(R) (2002).

\bibitem{romano2004}
  D. R. Romano and E. J. V. de Passos, Phys. Rev. A \textbf{70},
  043614 (2004).

\bibitem{zhang2005a}
  W. Zhang, D. L. Zhou, M.-S. Chang, M. S. Chapman, and L. You,
  Phys. Rev. A \textbf{72}, 013602 (2005).

\bibitem{saito2005}
  H. Saito and M. Ueda, Phys. Rev. A \textbf{72}, 053628 (2005).

\bibitem{luo2007}
  M. Luo, Z. Li, and C. Bao, Phys. Rev. A \textbf{75}, 043609 (2007).

\bibitem{xu2009}
  Z. F. Xu, Yunbo Zhang, and L. You, Phys. Rev. A \textbf{79}, 023613 (2009).

\bibitem{xu2010a}
  Z. F. Xu, Jie Zhang, Yunbo Zhang, and L. You, Phys. Rev. A \textbf{81}, 033603 (2010).

\bibitem{xu2010b}
  Jie Zhang, Z. F. Xu, L. You, and Yunbo Zhang, Phys. Rev. A \textbf{82}, 013625 (2010).

\bibitem{xu2010c}
  Z. F. Xu, J. W. Mei, R. L\"u, and L. You, Phys. Rev. A \textbf{82}, 053626 (2010).

\bibitem{shi2010}
  Yu Shi, Phys. Rev. A \textbf{82}, 023603 (2010).

\bibitem{xu2011}
  Z. F. Xu, R. L\"u, and L. You, Phys. Rev. A \textbf{84}, 063634 (2011).

\bibitem{stoof1988}
  H. T. C. Stoof, J. M. V. A. Koelman, and B. J. Verhaar, Phys. Rev. B \textbf{38},
  4688 (1988).

\bibitem{weiss2003}
  S. B. Weiss, M. Bhattacharya, and N. P. Bigelow, Phys. Rev. A \textbf{68},
  042708 (2003).

\bibitem{pashov2005}
  A. Pashov, O. Docenko, M. Tamanis, R. Ferber, H. Kn\"ockel, and E. Tiemann,
  Phys. Rev. A \textbf{72}, 062505 (2005).

\bibitem{Rb87}For $^{87}$Rb atoms£¬
$a_0 £½ 100.4 (a_B)$, $a_2 = 101.8 (a_B)$, as taken from
E. G. M. van Kempen, S. J. J. M. F. Kokkelmans, D. J. Heinzen and B. J. Verhaar,
Phys. Rev. Lett. {\bf 88}, 093201 (2002).

\bibitem{Na23}For $^{23}$Na atoms
$a_0 = 50.0 (a_B)$, $a_2 = 55.0 (a_B)$, as taken from
A. Crubellier, O. Dulieu, F. Masnou-Seeuws, M. Elbs, H. Knockel and
E. Tiemann, Eur. Phys. J. D {\bf 6}, 211 (1999).

\bibitem{gerbier2006}
  F. Gerbier, A. Widera, S. F\"olling, O. Mandel, and I. Bloch,
  Phys. Rev. A \textbf{73}, 041602(R) (2006).

\bibitem{leslie2009}
  S. R. Leslie, J. Guzman, M. Vengalattore, J. D. Sau, M. L. Cohen, and
  D. M. Stamper-Kurn, Phys. Rev. A \textbf{79}, 043631 (2009).

\end{thebibliography}
\end{document}